\documentclass[aps,preprint,showkeys]{revtex4-1}
\usepackage{graphicx}
\usepackage{amsmath}
\usepackage{makeidx}
\usepackage{amsfonts}
\usepackage{amssymb}
\usepackage{mathtools}
\usepackage{xcolor}

\newcommand{\ket}[1]{|{#1}\rangle}
\newcommand{\bra}[1]{\langle{#1}|}

\begin{document}

\title{Quantum Annealing with a Quantum Neural Network: Width of Annealing Matters}
\date{\today}

\author{Mojtaba Asadollahi}
%\email{mojtaba.77.ma@gmail.com}
\affiliation{Department of Physics, College of Science, Shiraz University, Shiraz 71454, Iran}

\author{Danial Khajehzadeh}
%\email{danialkhajehzadeh.2001@gmail.com}
\affiliation{Department of Physics, College of Science, Shiraz University, Shiraz 71454, Iran}

\author{Abolfazl Ramezanpour}
\email{aramezanpour@gmail.com}
\affiliation{Department of Physics, College of Science, Shiraz University, Shiraz 71454, Iran}
\affiliation{Leiden Academic Centre for Drug Research, Faculty of Mathematics and Natural Sciences, Leiden University, PO Box 9500-2300 RA Leiden, The Netherlands}

\begin{abstract}
Quantum annealing is a promising quantum algorithm for dealing with computationally hard optimization problems. Currently available hardware can implement the algorithm for a significant class of Hamiltonian problems. However, its performance is especially limited in discontinuous quantum phase transitions, where the energy gap shrinks exponentially with system size. In this paper, we introduce an extension of quantum annealing to a quantum neural network and study its performance in a number of prototypical optimization problems. In particular, we observe that already a random quantum annealing network of width $W=2$ can exponentially reduce the number of evolution layers or depth of the annealing process in the quantum search problem. By tuning the network parameters the search problem can be solved with exponentially small errors by a quantum perceptron of depth $L=1$ and width $W=2$. The findings show that width of a quantum annealing network is critical to enhance the success probability of the algorithm by providing a larger and learnable evolution space for the annealing process.  
\end{abstract}

\maketitle

\section{Introduction}\label{S0}
Quantum approaches have been shown to offer advantages across a range of computational problems \cite{Silva,Bravyi,Daley,Huang,Yamakawa}. Specifically, a variety of quantum and quantum-inspired methods have been developed to address computationally hard optimization problems \cite{Abbas}. Important examples include the quantum approximate optimization algorithm \cite{Farhi-qaoa-2014,Zhou-prx-2020} and quantum annealing (QA) algorithm \cite{Santoro,Morita}, among others \cite{Dupont,Cazalis,Alcazar,Fioroni}. In particular, quantum annealing has attracted considerable attention owing to its applicability to a broad class of problems \cite{Lucas}. In this approach, the objective function is encoded in the ground state of a problem Hamiltonian, which the system reaches through a controlled interpolation from an initial driver Hamiltonian whose ground state is easily prepared \cite{Farhi}.

Despite its appeal, quantum annealing faces several challenges that motivate further improvements \cite{Wang,Hauke-rpp-2020}. Most importantly, the energy gap between the ground state and the first excited state can decay exponentially with system size in case of first-order quantum phase transitions. As a result, the required running time of the algorithm grows exponentially with size, reducing the efficiency of the annealing process \cite{qxor-prl-2010,Young,qxor-pre-2011,qxor-pra-2012,QC-prep-2013,Brodoloni,Damerow-prb-2026}. On the other hand, coupling to the environment and the resulting decoherence can strongly disturb the adiabatic evolution \cite{Albash,Lidar-prl-2025}.

Several approaches have been proposed to address the above challenges. For instance, counterdiabatic driving and quantum optimal control improve the performance of quantum annealing by suppressing the destructive quantum transitions \cite{Muga-rmp-2019,Polkovnikov,Malla-prr-2024}. Forward and reverse annealing has been employed to increase the probability of finding the ground states \cite{Marshall,Mehta-pra-2025}. In addition, reinforcement learning can be used to optimize the annealing schedule based on previous runs \cite{Finin-srep-2020,ChenChen}. Variational annealing schedules that exploit Landau--Zener--St\"uckelberg (LZS) interference can yield exponential speedups over adiabatic quantum annealing for specific problems \cite{Werner-arxiv-2026}. These approaches either apply only to specific classes of problems, require additional measurements and optimization overhead, or do not fundamentally address robustness against decoherence.

\begin{figure}
\includegraphics[width=14cm]{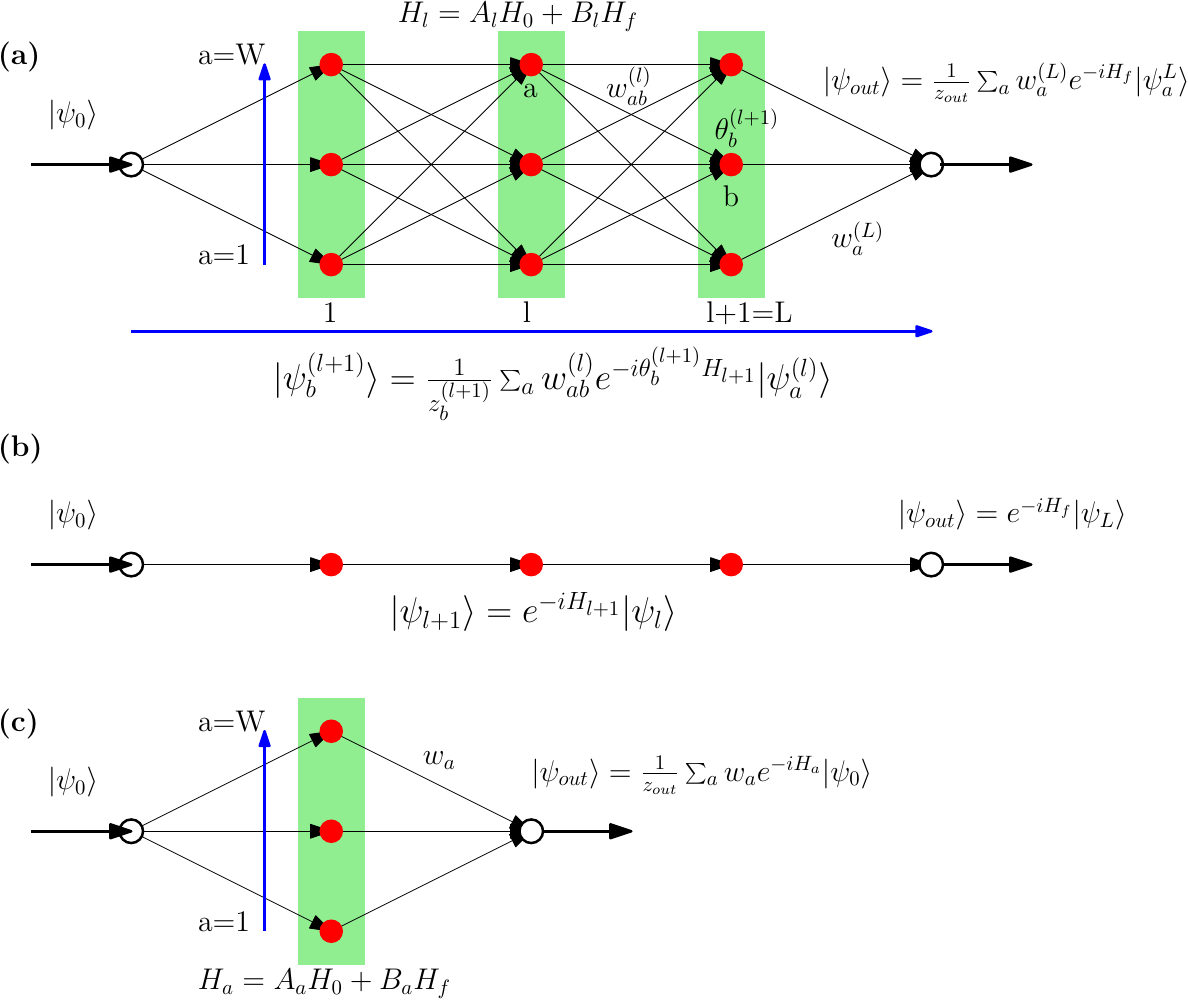} 
\caption{(a) Illustration of a quantum neural network of depth $L$ and width $W$ for quantum annealing. Depth $L$ denotes the number of annealing layers with width $W$. The coefficients $z_{b}^{(l+1)}$ are used for normalization of state. In a linear annealing $A_l^{(L)}=(1-l/L)$ and $B_l^{(L)}=l/L$. In the Grover annealing for the quantum search problem the coefficient $A_l^{(G)}$ and $B_l^{(G)}$ are given by the Grover algorithm. Panels (b) and (c) represent a standard quantum annealing and a quantum perceptron, respectively.}\label{fig0}
\end{figure}

In this paper, we extend the quantum annealing process to a quantum neural network (QNN) by introducing a width at each layer of the annealing dynamics. Figure \ref{fig0} shows an illustration of the proposed neural network. This architecture is intuitively promising as the annealing process now traverses multiple evolution paths, increasing the chance of getting around the exponentially small energy gaps. In addition, trainability of the neural network can help to find the possible optimal paths in the network for the annealing process. Starting from the ground state of the driver Hamiltonian, weighted sums of the states from the previous layer are fed as inputs to the nodes of the next layer. A unitary operation is then applied to the input state of each node, and this process is repeated for $L$ layers of width $W$ up to the single output node of the network. The Hamiltonians in the unitary evolutions change according to an annealing schedule, transitioning from the driver Hamiltonian in the first layer to the problem Hamiltonian in the last layer. In the following, we shall see  how the width of a quantum annealing process affects the performance of this algorithm in a number of well-studied optimization and search problems.

Note that the input of each node is a weighted sum of quantum states that are produced by different unitaries in the previous layers. The possibility of working with superpositions of unitary operators has recently attracted considerable attention \cite{Gui-comp-2006,Childs-arxiv-2012,Berry-prl-2014,Berry-ieee-2015,Qadd-prl-2016,Qadd-pra-2017}. Notably, it has been shown that such superpositions of evolutions can increase robustness against decoherence in a quantum dynamics \cite{Chatterjee-prl-2025}. Another strategy for speeding up annealing and increasing the success probability is to reinforce (steer) the quantum dynamics by exploiting information about the system's quantum state \cite{Ramezanpour2018,Ramezanpour2022,Homayouni}. This approach is however limited by the extensive resources required to evaluate the reinforcement terms. The above QNN achieves a speedup over standard quantum annealing by effectively modeling the reinforcement using the superposition of states from the previous layers, yet it does not suffer from the hardness introduced by reinforcement. Instead, here the complexity grows exponentially with depth as the number of states that are generated at the output node of the network is $W^L$.

The structure of this paper is as follows: In Sec. \ref{S1}, we present the problem statement and the precise structure of the QNN. Then, in Sec. \ref{S2}, we perform exact numerical simulations for the ferromagnetic p-spin model and quantum search problem. In Sec. \ref{S3}, we investigate the performance of approximate evolutions with a mean-field Hamiltonian in the random spin-glass model. The concluding remarks are given in Sec. \ref{S4}.

\section{Problem Statement}\label{S1}
Consider a quantum system of $N$ qubits in a Hilbert space of dimension $2^N$ with basis states $|\mathbf{x}\rangle$. Here  $\mathbf{x}=\{x_i=0,1,i=1\dots,N\}$ and the $|x_i\rangle$ are eigenstates of Pauli operators $\sigma_i^{z}$ with eigenvalues $\sigma_i=1-2x_i$.

The annealing process is approximated by $L$ layers of quantum evolution, as depicted in Fig.\ref{fig0}. The Hamiltonian in each layer $l=1,\dots,L$ is $H_l=A_lH_0+B_lH_f$. The Hamiltonian coefficients are $A_l=1-t_l$ and $B_l=t_l$, where $t_l\in (0,1)$ is a monotonically increasing function of layer index $l$. For instance, in a linear annealing $t_l=l/L$.

Each layer has width $W$, and at each node $b$ of layer $l+1$, we first compute the input state to the node
\begin{align}
|\alpha_b^{(l+1)}\rangle=\frac{1}{z_b^{(l+1)}}\sum_{a=1}^Ww_{ab}^{(l)}|\psi_a^{(l)}\rangle.
\end{align}
In Appendix \ref{app-US} we show how to obtain such a superposition given the unitaries that are needed to prepare the system states \cite{Chatterjee-prl-2025}.     
Then we use a coherent evolution represented by unitary $U_b^{(l+1)}=e^{-\mathrm{i}\theta_b^{(l+1)}H_{l+1}}$ to obtain
\begin{align}
|\psi_b^{(l+1)}\rangle=U_b^{(l+1)}|\alpha_b^{(l+1)}\rangle,
\end{align}
for $l=0,\dots,L-1$. The coefficient $z_b^{(l+1)}$ is for normalization to have $\langle \alpha_b^{(l+1)}|\alpha_b^{(l+1)}\rangle=1$. The network parameters are $\boldsymbol\theta=\{\theta_a^{(l)}: l=1,\dots,L; a=1,\dots,W\}$ and $\mathbf{w}=\{w_{ab}^{(l)}: l=1,\dots,L; a,b=1,\dots,W\}$.

The dynamics starts at layer $l=1$ with the ground state of $H_0$, which is denoted by the normalized state $|\psi_0\rangle$. So, the input state for a node in the first layer is $|\alpha_b^{(1)}\rangle=|\psi_0\rangle$.  
The output state after the last layer is normalized and is given by  
\begin{align}
|\psi_{out}\rangle=\frac{1}{z_{out}}\sum_{a=1}^Ww_{a}^{(L)}e^{-iH_f}|\psi_a^{(L)}\rangle.
\end{align}
The error function at the end is obtained from $E=\langle \psi_{out}|H_f|\psi_{out} \rangle$. The question is how much the width of the neural network for quantum annealing can enhance the performance of the annealing process.

Note that the standard quantum annealing (QA) process is recovered for $W=1$, where all the parameters $\theta_a^{(l)}=w_{ab}^{(l)}=1$. That is the parameters are the $A_l$ and $B_l$, which can be fixed according to an annealing schedule.  

The other limit is a quantum perceptron (QP) for $L=1$, where the output reads as follows
\begin{align}
|\psi_{out}\rangle=\frac{1}{z_{out}}\sum_{a=1}^W w_{a}e^{-iH_a}|\psi_0\rangle,
\end{align}
with $H_a=A_aH_0+B_aH_f$. For the quantum perceptron, we take the $A_a$ and $B_a$ as free parameters along with the weights $w_a$.

\section{Exact numerical simulations}\label{S2}

\subsection{Ferromagnetic p-spin Model}\label{S21}
Let us start with the computationally easy problem of finding the ground state of a ferromagnetic and fully-connected p-spin model. For $p=3$ the problem Hamiltonian reads 
\begin{align}
H_f=-N(\frac{1}{N}\sum_{i=1}^N\sigma_i^z)^3.
\end{align}
The ground state is trivial with all spins up and energy $E_g=-N$.
The initial Hamiltonian is $H_0=-\sum_{i=1}^N\sigma_i^x$. This system displays a discontinuous quantum phase transition in a quantum annealing process around $t_l\sim 0.44$ where the energy gap becomes exponentially small in system size \cite{Bapst-jstat-2012}. Consequently, the annealing time should be exponentially large to reach the trivial ground state of the problem starting from that of $H_0$.

\subsubsection{The Quantum Perceptron}\label{S211}
\begin{figure}
	\includegraphics[width=14cm]{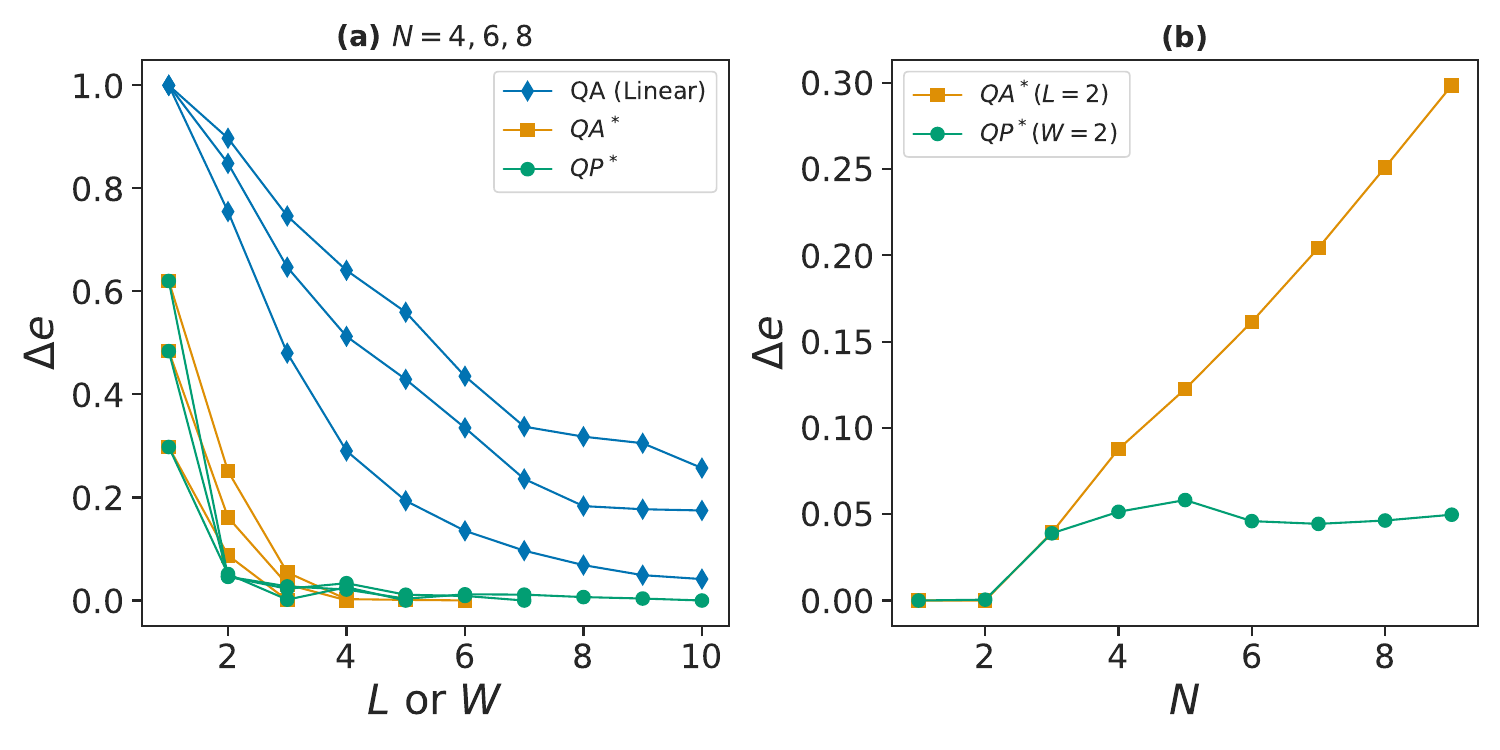} 
	\caption{Relative error of an optimal quantum perceptron ($QP^*$) in the ferromagnetic p-spin model. It is compared with the optimal quantum annealing ($QA^*$) and the standard quantum annealing ($QA$) with linear annealing schedule. Here $p=3$ and the relative error is $\Delta e=(\langle H_f\rangle+N)/N$. Panel (a) shows the error vs depth $L$ (in $QA$ and $QA^*$) or width $W$ (in $QP^*$). Panel (b) shows the error vs the system size $N$ for $L=2$ ($W=2$). We used the "differential\_evolution" (popsize=$50$, maxiter=$100$) and "minimize" (SLSQP) methods of "scipy.optimize" to find the optimal parameters. All the parameters are limited to $(-1,+1)$.}\label{fig1}
\end{figure}

Here we compare the performance of the standard quantum annealing (linear schedule) and the optimal quantum annealing with that of the optimal quantum perceptron. In an optimal annealing of depth $L$, the parameters $A_l,B_l \in (-1,+1)$ are optimized to minimize the expected energy function $E$. Similarly, in an optimal perceptron of width $W$ we find the optimal parameters $A_a,B_a,w_a \in (-1,+1)$ which minimize $E=\langle \psi_{out}|H_f|\psi_{out} \rangle$. Figure \ref{fig1} shows the relative error $\Delta e=(E-E_g)/N$ vs depth $L$ (in QA) or width $W$ (in QP) for small problem sizes $N$. The optimal QA obviously performs much better than the standard QA with errors that are comparable to those of the optimal perceptron. Nevertheless, as the figure (panel (b)) shows, already for width $W=2$ the perceptron works very well compared to the optimal QA of depth $L=2$. At least for the reported problem sizes $N<10$, the perceptron error grows very slowly with $N$ unlike the optimal QA. This observation indicates the high impact of width in the algorithm relative to the depth of the network.         

In the following, we consider a computationally more challenging problem to study the effect of width of the quantum annealing neural network on its performance. 

\subsection{Quantum Search Problem}\label{S22}
The quantum search problem can be formulated as a quantum annealing process from the ground state
of $H_0=\mathbb{I}-|\psi_0\rangle\langle \psi_0|$ to that of $H_f=\mathbb{I}-|\psi_f\rangle\langle \psi_f|$. Here $|\psi_f\rangle$ is the target state and 
\begin{align}
	|\psi_0\rangle=\frac{1}{\sqrt{2^N}}\sum_{\boldsymbol\sigma}|\boldsymbol\sigma\rangle=\sqrt{P_0}|\psi_f\rangle+\sqrt{1-P_0}|\psi_f^{\perp}\rangle.
\end{align}
We use $|\psi_f^{\perp}\rangle$ to represent the subspace of states that are orthogonal to $|\psi_f\rangle$.
$P_0=1/2^N$ is the probability of finding the target state in $|\psi_0\rangle$.

The optimal annealing schedule is given by the Grover algorithm \cite{grover-prl-1997,Roland-pra-2002,fdsearch-prl-2005,fdsearch-pra-2017}
\begin{align}\label{grover}
	t_l^{(G)} &=\frac{1}{2}\left[1-\sqrt{\frac{P_0}{1-P_0}}\tan\left((1-2\frac{l}{L})\phi\right)\right],\\
	\phi &=\arctan\left(\sqrt{\frac{1-P_0}{P_0}}\right).
\end{align}
In this case, the Hamiltonian coefficients are denoted by $A_l^{(G)}=1-t_l^{(G)}, B_l^{(G)}=t_l^{(G)}$.
The Grover scaling $L_{min}^{(G)}\propto \sqrt{2^N}$ gives the minimum number of evolution layers $L$ which is needed to have a success probability $P_{success}=|\langle \psi_{out}|\psi_f\rangle|^2$ that approaches $1$ for large $N$. The scaling is optimal for linear-oracle model of unstructured search problem, where no other information about the problem is provided.

\subsubsection{The Quantum Perceptron}\label{S221}
Let us start with the quantum perceptron ($L=1$). Here we show that even for width $W=2$ the error approaches zero with system size $N$. The output state of the perceptron reads as follows
\begin{align}
		|\psi_{out}\rangle =\frac{1}{z_{out}}\sum_{a=1}^{W} w_a e^{-i(A_a+B_a)/2}\left(C_a|\psi_f\rangle+C_a^{\perp}|\psi_f^{\perp}\rangle \right),
\end{align}
with
\begin{align}
		C_a &= \sqrt{P_0}\cos(n_a) + i\frac{\sin(n_a)}{n_a}\left( \sqrt{1-P_0}n_a^x + \sqrt{P_0}n_a^z\right),\\
		C_a^{\perp} &=\sqrt{1-P_0}\cos(n_a) + i \frac{\sin(n_a)}{n_a} \left(\sqrt{P_0}n_a^x - \sqrt{1-P_0}n_a^z\right),
\end{align}
and
\begin{align}
	n_a^x &= A_a\sqrt{P_0(1-P_0)},\\
	n_a^z &=\frac{2A_aP_0-A_a+ B_a}{2},\\
	n_a &= \sqrt{({n_a^x})^2 + ({n_a^z})^2}.
\end{align}

In Table \ref{tab-1} we report the optimal parameters for different values of $P_0$ when $W=2$.

\begin{table}[t]
	\centering
	\begin{tabular}{|l|c|c|c|c|c|c|}
	    \hline
		$P_0$ & $w_1$ & $w_2$ & $A_1$ & $A_2$ & $B_1$ & $B_2$ \\
	    \hline
		$1/2^{4}$    &  0.956 & -0.934 &  0.169 & -0.962 &  0.829 &  0.843 \\
	    \hline
		$1/2^{8}$   &  0.977 & -0.979 &  0.707 & -0.805 & -0.931 & -0.931 \\
	    \hline
		$1/2^{16}$ & -0.828 &  0.828 & -0.659 &  0.624 & -0.739 & -0.739 \\
	    \hline
	\end{tabular}
	\caption{Optimal parameters of the quantum perceptron ($L=1$) with width $W=2$ for different values of $P_0$ in the quantum search problem. We used the "differential\_evolution" (popsize=$50$, maxiter=$100$) and "minimize" (SLSQP) methods of "scipy.optimize" to find the optimal solutions. The parameters are limited to $(-1,+1)$.}\label{tab-1}
\end{table}

The data in the table suggest the parameter values $w_1=-w_2=1, A_1=-A_2=1,B_1=B_2=1$ for $P_0\to 0$. Using these parameters, we obtain
\begin{align}
	|\psi_{out}\rangle &= \frac{1}{z_{out}}(C|\psi_f\rangle+C^{\perp}|\psi_f^{\perp}\rangle),\\
	C &= e^{-i}\sqrt{P_0}\cos\sqrt{P_0} + e^{-i} i\sin\sqrt{P_0}-\sqrt{P_0}\cos\sqrt{1-P_0},\\ 
	C^{\perp} &= e^{-i}\sqrt{1-P_0}\cos\sqrt{P_0}-\sqrt{1-P_0}\cos\sqrt{1-P_0}+i\sin\sqrt{1-P_0}.
\end{align}
Expanding for $P_0\to 0$, we find $C\propto \sqrt{P_0}$ and $C^{\perp}\propto P_0$. Therefore, to leading order in $P_0$, the error probability is by a factor $\sqrt{P_0}$ smaller than the success probability. As a result, the error approaches zero with $P_0$, considering the normalization factor. Note that, in contrast to the Grover algorithm, the quantum evolution here is not linear.

In the following, we see how a random neural network with an annealing schedule, given by the parameters $A_l,B_l$, performs without adjusting the network parameters.

\subsubsection{Quantum annealing with a random neural network}\label{S222}
\begin{figure}
\includegraphics[width=14cm]{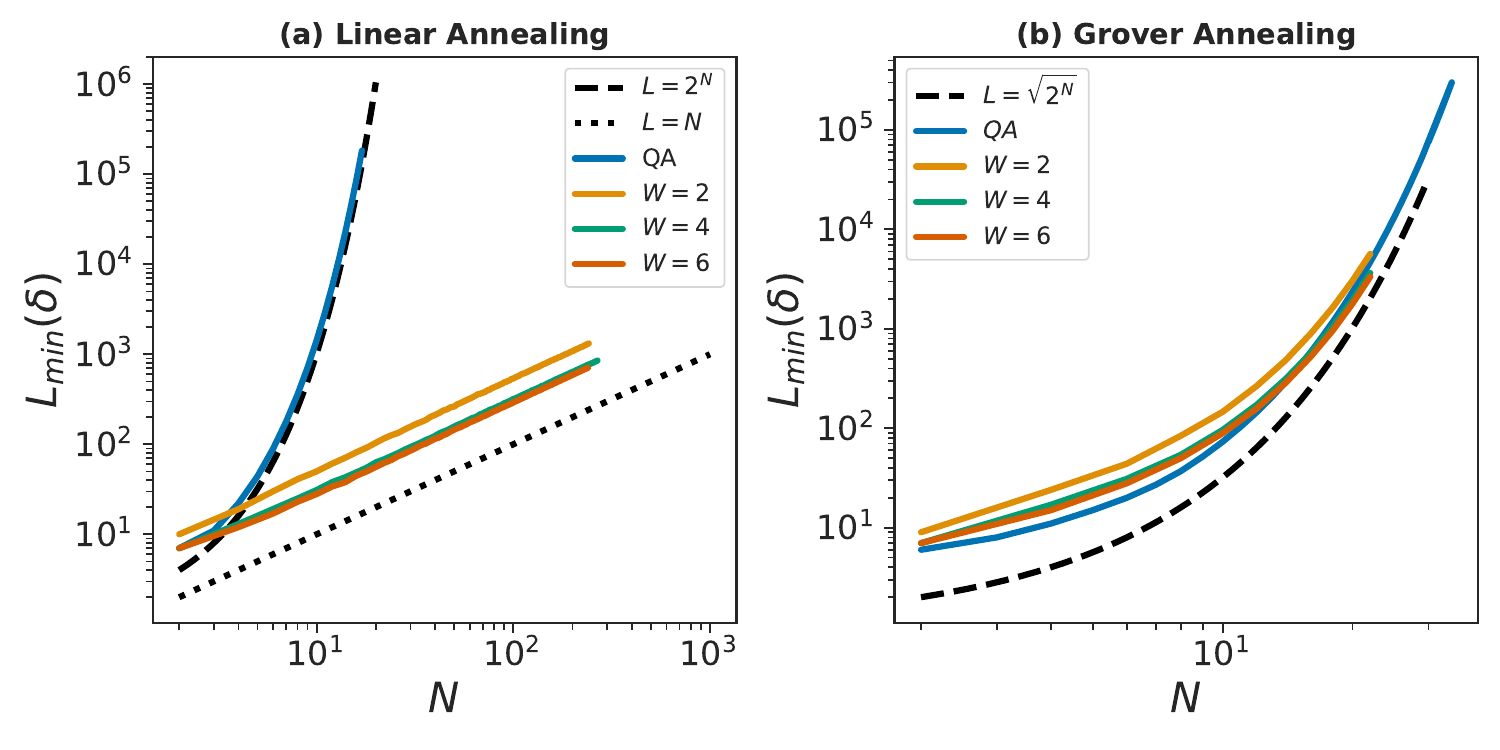} 
\caption{Minimal depth $L_{min}(\delta)$ of a random quantum annealing network in the quantum search problem. Here $L_{min}(\delta)$ is the minimum depth where the average success probability becomes greater than $1-\delta$. Here $\delta=1/3$. Panels (a) and (b) compare the performance of standard quantum annealing (QA) for linear and Grover annealing schedules with that of the quantum neural network for different width $W$. The weights $w_{ab}^{(l)}$ of neural network are independent random variables uniformly distributed in $(0,1)$. The phases $\theta_a^{(l)}$ are independent random variables from a normal distribution of mean zero and variance $1$.}\label{fig2}
\end{figure}

We consider a random neural network where the parameters $\theta_a^{(l)}\sim \mathcal{N}(0,1)$ are independent random variables from a normal distribution of mean zero and variance one. The $w_{ab}^{(l)}$ and $w_a^{(L)}$ are also independent random variables uniformly distributed in $(0,1)$. Figure \ref{fig2} shows the minimum number of layers $L_{min}(\delta)$ where the average error $E=1-P_{success}$ becomes less than $\delta=1/3$ for linear and Grover annealing schedules. We see that even a random network with linear annealing can reduce the scaling of $L_{min}$ from $\sqrt{2^N}$ to $N$. In this case, a constructive superposition of unitary evolutions helps to exponentially reduce the number of layers (depth) of the quantum annealing network. Unexpectedly, the width of networks is not relevant in the Grover case where the annealing is very slow around the quantum phase transition $A_l=B_l=1/2$. It seems that a linear annealing of width $W$ provides a larger number of dynamical paths to the annealing process than the Grover annealing; in the latter case, a large number of layers remain nearly identical near the phase transition and therefore do not contribute to expanding the evolution space.

To study larger problem sizes, in the next section we consider an approximate version of the quantum annealing neural network.

\section{Approximate evolutions with mean-field Hamiltonians}\label{S3}
Here we present a quantum neural network that uses a mean-field approximation of the problem Hamiltonian. The mean-field parameters can be optimized to minimize the error function along with the other parameters of the network $\{\boldsymbol\theta,\mathbf{w}\}$. 

More precisely, in any node we use a mean-field approximation of the problem Hamiltonian
\begin{align}
H_f^{(MF)}=-\sum_i J_i\sigma_i^z.
\end{align}
This allows us to write the unitary $U_a^{(l)}$ in a factorized form
\begin{align}
U_a^{(l)} &=\prod_i e^{-\hat{i}\theta_a^{(l)}H_i^{(l)}},\\
H_i^{(l)} &=-A_l\sigma_i^x-B_lJ_i\sigma_i^z.
\end{align}

The dynamics begins with a product state $\ket{\psi_0}=\prod_{i=1}^N\otimes\ket{+}_i$ where $|+\rangle_i=(|0\rangle_i+|1\rangle_i)/\sqrt{2}$. Therefore the output state $\ket{\psi_a^{(l)}}$ of node $a$ at layer $l$ is a superposition of $W^{l-1}$ normalized product state. At the end, the loss function is computed by the original problem Hamiltonian $E = \bra{ \psi_{out}} H_f \ket{ \psi_{out}}$. More details are given in Appendix \ref{app-MF}.

\begin{figure}
	\includegraphics[width=14cm]{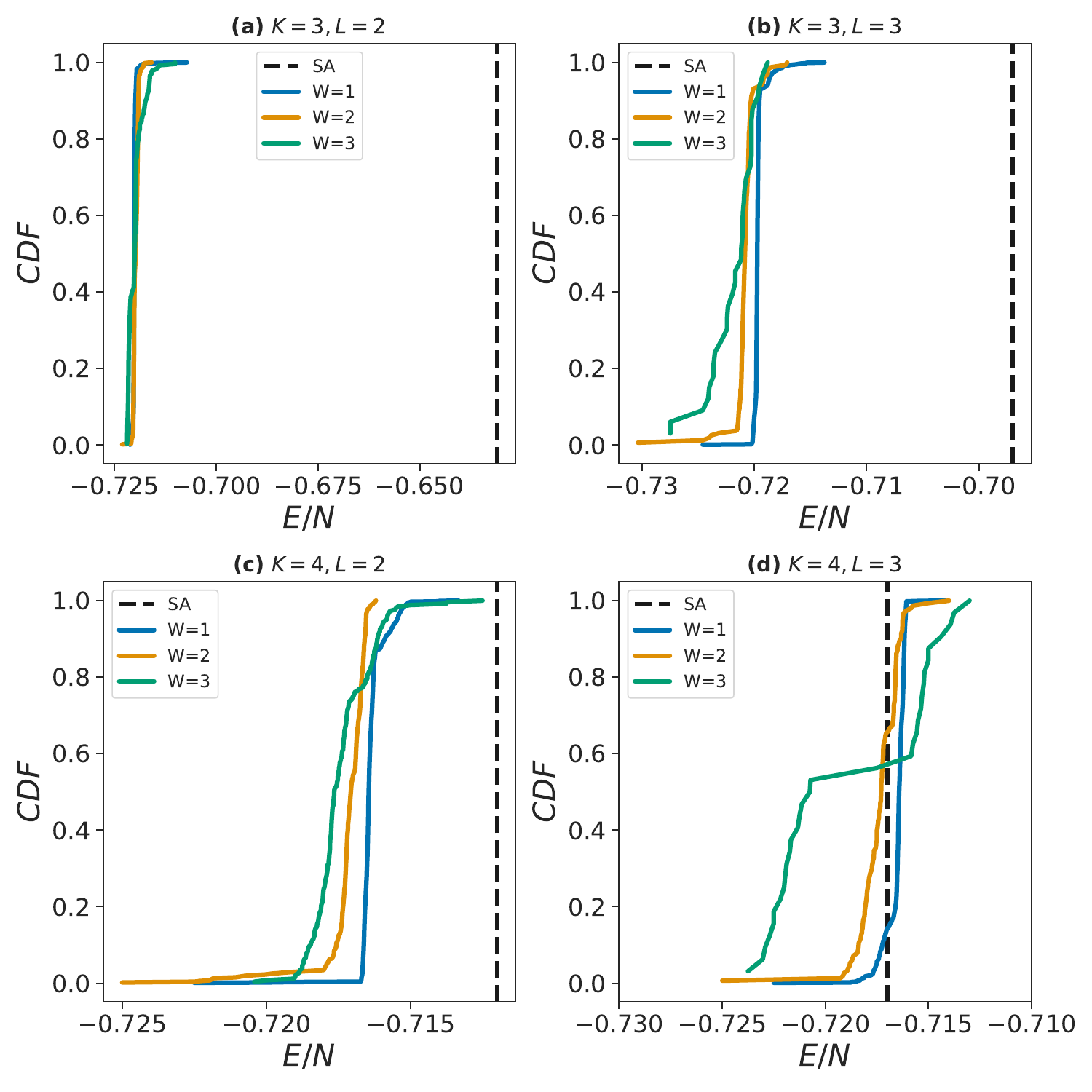} 
	\caption{Cumulative distribution of minimum energies found by the mean-field random neural network in the random spin-glass model. The results are compared with the average energy obtained with the standard simulated annealing ($SA$) for the same number of layers and iterations (Monte Carlo sweeps) that is $T=1000$. The spin-glass model is defined on a random regular graph of degree $K$ with the random couplings $J_{ij}=\pm 1/\sqrt{K}$. Panels (a) and (b) show the results for different depths $L=2,3$ and widths $W=1,2,3$ with system size $N=400$ and degree $K=3$. Panels (c) and (d) show the results for different depths $L=2,3$ and widths $W=1,2,3$ with system size $N=400$ and degree $K=4$. The number of independent realizations changes from at least $30$ for $L=3, W=3$ to $1000$ for $L=2,W=1$.}\label{fig3}
\end{figure}

\subsection{Random Spin-Glass Model}\label{S31}
We consider a spin-glass model on a random regular graph $G$ of degree $K$ where each Ising variable $\sigma_i=\pm 1$ interacts with $K$ other randomly selected variables. The set of neighboring nodes of spin $i$ are denoted by $\partial i$. The strength of two-spin interactions are independent and random variables $J_{ij}=\pm 1$ that take the two values with equal probability. The problem Hamiltonian is
\begin{align}
H_f=-\sum_{(ij)\in G} \frac{J_{ij}}{\sqrt{K}}\sigma_i^z\sigma_j^z.
\end{align}

Again we take a random neural network and minimize the error function by updating only the mean-field parameters $J_i$. The other parameters $\boldsymbol\theta$ and $\mathbf{w}$ are independent random variables where as before $\theta_a^{(l)}\sim \mathcal{N}(0,1)$ and $w_{ab}^{(l)}, w_a^{L}$ are uniformly distributed in $(0,1)$.
The mean-field parameters $J_i$ are updated to find self-consistent fields resulting to smaller energy values. We start with initially random fields $J_i(0)\in (-1,1)$. In each step $t$ we run the forward algorithm to compute the average magnetization $m_i(t)=\langle \psi_{out}|\sigma_i^z|\psi_{out}\rangle$. The candidate solution and the fields are obtained from $s_i(t)=\mathrm{sign}(m_i(t))$ and $J_i(t)=\sum_{j\in \partial_i}J_{ij}s_j/2$, respectively. In the next step the fields at each node are replaced with
\begin{align}
J_i(t+1)=(1-\lambda)J_i(t)+\lambda J_i^{opt}(t)+\eta_i(t),
\end{align}
where $\mathbf{J}^{opt}(t)$ is the optimal field up to time step $t$. The added noise $\eta_i$ is a normal variable of mean zero and variance $\Delta^2(t)$. Here $\Delta(t)$ decreases with time from $1$ at time step $t=0$ to $e^{-2}$ at the maximum time step $T=1000$, and we set $\lambda=0.1$. The parameters have roughly been adjusted to obtain a good performance.

Figure \ref{fig3} displays the cumulative distribution of minimum energies we find in a number of independent realizations of the problem Hamiltonian with $N=400$ qubits. The number of realizations varies from at least $30$ for the case $L=3, W=3$ to $1000$ for $L=2, W=1$. As the figure shows, the probability of finding smaller energies is enhanced for larger depth and width even for these random networks. For comparison, we also show the average energy that is obtained by the standard simulated annealing (SA) algorithm where the inverse temperature increases linearly $\beta_l=(l/L)\beta_{max}$ with layer index $l=1,\dots,L$. Here $\beta_{max}=10$ and the total number of Monte Carlo sweeps is $T$ to have a fair comparison. We see that even a mean-field approximation with a random neural network can perform comparably to or better than the SA algorithm, with the difference being more pronounced for small depth $L=2$.

\section{Conclusion}\label{S4}
In summary, we presented a quantum annealing neural network and observed that the width of network can significantly enhance the annealing performance in some optimization and search problems. In the ferromagnetic p-spin model, a quantum perceptron of depth $L=1$ and width $W=2$ results in negligible errors that do not increase with problem size for $N<10$, indicating that the effects of the exponentially small energy gap are being mitigated in this problem. Moreover, an optimized quantum perceptron (depth $L=1$) already gives exponentially small errors for width $W=2$ in the quantum search problem. Here, a random quantum annealing network exponentially reduces the minimal depth of the network from $\sqrt{2^N}$ to $N$. We also presented a mean-field approximation of the quantum annealing network for classical simulation of the annealing process with larger problem sizes. In this case again a random neural network performs better than or comparably to standard simulated annealing algorithm in a random spin-glass model.

As mentioned in the main text, complexity of the quantum annealing network grows exponentially with depth as $W^{L}$. This is the number of states we need to reconstruct at the output node of the network. The required unitaries are provided by the network structure and parameters which are of order $L(W^2+W)$. Each unitary can be implemented by about $N/\epsilon$ local gates for a local and sparse problem Hamiltonian using the Trotter approximation with an error of order $\epsilon^2$. In addition, the success probability of obtaining the right superposition is proportional to $1/W^{L}$, in the worst case of orthogonal states. All together, it means that to have a reasonable complexity the depth and width of the network should grow slower than $N$ with system size. This highlights the role of learning and optimizing the network parameters as we observed in the case of quantum perceptron in the quantum search problem.

Finally, we need to go beyond the mean-field approximation and use more powerful learning algorithms for classical simulation of the quantum annealing neural network in larger systems. And, we are working on a similar extension of the classical simulated annealing algorithm to a neural network of simulated annealing. As in the quantum case, we expect the width of network to be useful for mitigating the destructive effects of classical phase transitions by enlarging the search space of the annealing process.

\acknowledgments
We would like to thank Mohammad Hossein Zarei for helpful discussions.
This work was performed using the ALICE compute resources provided by Leiden University.

%%%%%%%%%%%%%%%%%%%%%%%%%%%

\clearpage
\newpage

%%%%%%%%%%%%%%%%%%%%%%%%%%%
\clearpage
\newpage

\appendix
\counterwithin{figure}{section}

\section{Linear Combination of Unitaries via Ancilla + Post-Selection}\label{app-US}
In this section we see how a superposition of system sates $\ket{\psi_i}$ can be obtained from the associated unitaries \cite{Gui-comp-2006,Childs-arxiv-2012,Berry-prl-2014}. Suppose we are to prepare the following target state
\begin{equation}
	\ket{\Psi} = \frac{\sum_{i=1}^{n} w_i\ket{\psi_i}}
	{\big\|\sum_{i=1}^{n} w_i \ket{\psi_i}\big\|},
	\qquad
	\sum_{i=1}^{n} |w_i|^2 = 1,
\end{equation}
from $n$ states $\ket{\psi_i}=V_i\ket{0}_S$ defined by the unitaries $V_i$.  Here $\big\|\ket{\psi}\big\|$ denotes the norm of state $\ket{\psi}$. At each node of layer $l$ the number of such states is $n=W^{l-1}$ for a network of width $W$. In our problem the unitaries $V_i$ are completely defined by $l-1$ phase parameters and the annealing parameters.  

Let $n=2^k$, and consider $k$ ancilla qubits. Apply a state-preparation unitary $F$ on the ancilla,
\begin{equation}
	F \ket{0}^{\otimes k} = \sum_{i=1}^{n} w_i \ket{i}_a ,
\end{equation}
and then the controlled preparation $\sum_{i=1}^{n}\ket{i}\bra{i}_a\otimes V_i$ on $\sum_{i=1}^{n} w_i \ket{i}_a\otimes\ket{0}_S$,
so that the joint state is
\begin{equation}
	\sum_{i=1}^{n} w_i \ket{i}_a \ket{\psi_i}.
\end{equation}
Project the ancilla onto
\begin{equation}
	\ket{\chi}_a = \frac{1}{\sqrt{n}} \sum_{i=1}^{n} \ket{i}_a.
\end{equation}
Then
\begin{equation}
	\bra{\chi}_a
	\Big[ \sum_{i=1}^{n} w_i \ket{i}_a \ket{\psi_i} \Big]
	= \frac{1}{\sqrt{n}} \sum_{i=1}^{n} w_i \, \ket{\psi_i},
\end{equation}
and after normalization we get
\begin{equation}
		\ket{\Psi} = \frac{\sum_{i=1}^{n} w_i \ket{\psi_i}}
		{\big\|\sum_{i=1}^{n} w_i \ket{\psi_i}\big\|}.
\end{equation}
The success probability for this outcome is
\begin{equation}
	p = \frac{1}{n}\Big\|\sum_{i=1}^{n} w_i \ket{\psi_i}\Big\|^2 ,
\end{equation}
which equals $1/n$ when the states $\ket{\psi_i}$ are orthonormal.

The same construction is applied independently to every node of layer $l$, each with its own ancilla and its own weights. Since the circuits $V_i$ are known, the branch states are regenerated for each node rather than copied. Therefore, this is not the cloning of an unknown quantum state.

\section{Quantum annealing neural networks with factorized unitaries}\label{app-MF}
In any intermediate node we use a mean-field approximation of the problem Hamiltonian
\begin{align}
H_f^{(MF)}=-\sum_i J_i\sigma_i^z.
\end{align}
This allows us to write the unitary $U_a^{(l)}$ in a factorized form
\begin{align}
U_a^{(l)}=\prod_i e^{-\hat{i}\theta_a^{(l)}H_i^{(l)}},\\
H_i^{(l)}=-A_l\sigma_i^x-B_lJ_i\sigma_i^z,\\
U_a^{(l)}=\prod_i\left(\cos(\theta_a^{(l)}\Omega_i^{(l)})-\hat{i}\frac{\sin(\theta_a^{(l)}\Omega_i^{(l)})}{\Omega_i^{(l)}}H_i^{(l)}\right),\\
\Omega_i^{(l)}=\sqrt{(A_l)^2+(B_lJ_i)^2}.
\end{align}
The mean-field parameters $J_i$ are to reduce the error like the other network parameters. 

The initial state is factorized $\ket{\psi_0}=\prod_{i=1}^N\otimes\ket{+}_i$ where $|+\rangle_i=(|0\rangle_i+|1\rangle_i)/\sqrt{2}$. Therefore, the output state $\ket{\psi_a^{(l)}}$ of node $a$ at layer $l$ is a superposition of $n_l=W^{l-1}$ normalized product states.

The loss function is computed with the original problem Hamiltonian:
\begin{align}
	E = \bra{ \psi_{out}} H_f \ket{ \psi_{out}},
\end{align}
where now
\begin{align}
|\psi_{out}\rangle=\frac{1}{z_{out}}\sum_{a=1}^Ww_{a}^{(L)}(\frac{1}{z_a^{(L)}}\sum_{n=1}^{n_L}C_{an}^{(L)}|\mathbf{h}_{an}^{(out)}\rangle),
\end{align}
with
\begin{align}
|\mathbf{h}_{an}^{(out)}\rangle=e^{-\hat{i}H_f^{(MF)}}|\mathbf{h}_{an}^{(L)}\rangle.
\end{align}

\subsection{Computing the product states}
Given the states of layer $l$
\begin{align}
	\ket{\psi_a^{(l)}} &=\frac{1}{z_{a}^{(l)}}\sum_{n=1}^{n_{l}} C_{an}^{(l)}\ket{\mathbf{h}_{an}^{(l)}},\\
	\ket{\mathbf{h}_{an}^{(l)}} &=\prod_{i=1}^N\otimes \left(\frac{e^{h_{an,i}^{(l)}\sigma_i^z}}{\sqrt{\cosh(h_{an,i}^{(l)}+h_{an,i}^{(l)*})}}|+\rangle_i\right),
\end{align}
the output of node $b$ at layer $l+1$ reads as follows
\begin{multline}
	\ket{\psi_b^{(l+1)}} =\frac{1}{z_{b}^{(l+1)}}\sum_a w_{ab}^{(l)}U_b^{(l+1)}\ket{\psi_a^{(l)}}
	=\frac{1}{z_{b}^{(l+1)}}\sum_{a=1}^W\sum_{n=1}^{n_l} \frac{1}{z_a^{(l)}}w_{ab}^{(l)}C_{an}^{(l)}U_b^{(l+1)}\ket{\mathbf{h}_{an}^{(l)}}\\
	=\frac{1}{z_{b}^{(l+1)}}\sum_{m=1}^{n_{l+1}} C_{bm}^{(l+1)}\ket{\mathbf{h}_{bm}^{(l+1)}},
\end{multline}
with $m=(an)$ and
\begin{align}
C_{bm}^{(l+1)} &=\frac{1}{z_a^{(l)}}w_{ab}^{(l)}C_{an}^{(l)},\\
\ket{\mathbf{h}_{bm}^{(l+1)}} &=U_b^{(l+1)}\ket{\mathbf{h}_{an}^{(l)}}.
\end{align}
In terms of the components we have
\begin{align}
h_{bm,i}^{(l+1)} =\frac{1}{2}\ln \left(\frac{U_{b,i}^{(l+1)}(0,0)e^{h_{an,i}^{(l)}}+U_{b,i}^{(l+1)}(0,1)e^{-h_{an,i}^{(l)}}}{U_{b,i}^{(l+1)}(1,0)e^{h_{an,i}^{(l)}}+U_{b,i}^{(l+1)}(1,1)e^{-h_{an,i}^{(l)}}}\right),
\end{align}
where
\begin{align}
U_{a,i}^{(l)}(0,0) &=\cos(\theta_a^{(l)}\Omega_i^{(l)})+\hat{i}\frac{\sin(\theta_a^{(l)}\Omega_i^{(l)})}{\Omega_i^{(l)}}B_lJ_i,\\
U_{a,i}^{(l)}(0,1) &=\hat{i}\frac{\sin(\theta_a^{(l)}\Omega_i^{(l)})}{\Omega_i^{(l)}}A_l,\\
U_{a,i}^{(l)}(1,0) &=\hat{i}\frac{\sin(\theta_a^{(l)}\Omega_i^{(l)})}{\Omega_i^{(l)}}A_l,\\
U_{a,i}^{(l)}(1,1) &=\cos(\theta_a^{(l)}\Omega_i^{(l)})-\hat{i}\frac{\sin(\theta_a^{(l)}\Omega_i^{(l)})}{\Omega_i^{(l)}}B_lJ_i.
\end{align}

\subsection{Computing the loss function}
We have
\begin{align}
	E =	\frac{1}{z_{out}^2}\sum_{a,b=1}^{W}\sum_{m,n=1}^{n_L}\frac{w_a^{(L)}C_{an}^{(L)}}{z_a^{(L)}}\frac{w_b^{(L)}C_{bm}^{(L)*}}{z_b^{(L)}}\langle \mathbf{h}_{bm}^{(out)}|H_f|\mathbf{h}_{an}^{(out)}\rangle,
\end{align}
where
\begin{align}
	z_{out}^2=\sum_{a,b}\sum_{m,n}\frac{w_a^{(L)}C_{an}^{(L)}}{z_a^{(L)}}\frac{w_b^{(L)}C_{bm}^{(L)*}}{z_b^{(L)}}\langle \mathbf{h}_{bm}^{(L)}|\mathbf{h}_{an}^{(L)}\rangle,\\
	(z_{a}^{(L)})^2=\sum_{m,n}C_{an}^{(L)}C_{am}^{(L)*}\langle \mathbf{h}_{am}^{(L)}|\mathbf{h}_{an}^{(L)}\rangle.
\end{align}
Here
\begin{align}
	\langle \mathbf{h}_{bm}^{(L)}|\mathbf{h}_{an}^{(L)}\rangle=\prod_i\left(\frac{\cosh(h_{an,i}^{(L)}+h_{bm,i}^{(L)*})}{\sqrt{\cosh(2R_{an,i}^{(L)})\cosh(2R_{bm,i}^{(L)})}} \right),
\end{align}
and
\begin{align}
	\frac{1}{\langle \mathbf{h}_{bm}^{(out)}|\mathbf{h}_{an}^{(out)}\rangle}\langle \mathbf{h}_{bm}^{(out)}|H_f|\mathbf{h}_{an}^{(out)}\rangle=-\sum_{(ij)\in G}\frac{J_{ij}}{\sqrt{K}}
\tanh(h_{an,i}^{(out)}+h_{bm,i}^{(out)*})\tanh(h_{an,j}^{(out)}+h_{bm,j}^{(out)*}).
\end{align}
We use $\mathbf{R}_{bm}^{(l+1)}$ and $\mathbf{I}_{bm}^{(l+1)}$ for the real and imaginary parts of the complex variables $\mathbf{h}_{bm}^{(l+1)}$. We assume that all parameters $\theta_{a}^{(l)}$, $w_{ab}^{(l)}$, and $J_i$ are real.


\begin{thebibliography}{prsty}

\bibitem{Silva} Ristè, D., Da Silva, M.P., Ryan, C.A., Cross, A.W., Córcoles, A.D., Smolin, J.A., Gambetta, J.M., Chow, J.M. and Johnson, B.R., 2017. Demonstration of quantum advantage in machine learning. npj Quantum Information, 3(1), p.16.

\bibitem{Bravyi} Bravyi, S., Gosset, D. and König, R., 2018. Quantum advantage with shallow circuits. Science, 362(6412), pp.308-311.

\bibitem{Daley} Daley, A.J., Bloch, I., Kokail, C., Flannigan, S., Pearson, N., Troyer, M. and Zoller, P., 2022. Practical quantum advantage in quantum simulation. Nature, 607(7920), pp.667-676.

\bibitem{Huang} Huang, H.Y., Broughton, M., Cotler, J., Chen, S., Li, J., Mohseni, M., Neven, H., Babbush, R., Kueng, R., Preskill, J. and McClean, J.R., 2022. Quantum advantage in learning from experiments. Science, 376(6598), pp.1182-1186.

\bibitem{Yamakawa} Yamakawa, T. and Zhandry, M., 2024. Verifiable quantum advantage without structure. Journal of the ACM, 71(3), pp.1-50.



\bibitem{Abbas} Abbas, A., Ambainis, A., Augustino, B., Bärtschi, A., Buhrman, H., Coffrin, C., Cortiana, G., Dunjko, V., Egger, D.J., Elmegreen, B.G. and Franco, N., 2024. Challenges and opportunities in quantum optimization. Nature Reviews Physics, 6(12), pp.718-735.



\bibitem{Farhi-qaoa-2014} Farhi, E., Goldstone, J. and Gutmann, S., 2014. A quantum approximate optimization algorithm. arXiv preprint arXiv:1411.4028.

\bibitem{Zhou-prx-2020} Zhou, L., Wang, S.T., Choi, S., Pichler, H. and Lukin, M.D., 2020. Quantum approximate optimization algorithm: Performance, mechanism, and implementation on near-term devices. Physical Review X, 10(2), p.021067.


\bibitem{Santoro} Santoro, G.E. and Tosatti, E., 2006. Optimization using quantum mechanics: quantum annealing through adiabatic evolution. Journal of Physics A: Mathematical and General, 39(36), pp.R393-R431.

\bibitem{Morita} Morita, S. and Nishimori, H., 2008. Mathematical foundation of quantum annealing. Journal of Mathematical Physics, 49(12).



\bibitem{Dupont} Dupont, M., Evert, B., Hodson, M.J., Sundar, B., Jeffrey, S., Yamaguchi, Y., Feng, D., Maciejewski, F.B., Hadfield, S., Alam, M.S. and Wang, Z., 2023. Quantum-enhanced greedy combinatorial optimization solver. Science Advances, 9(45), p.eadi0487.

\bibitem{Cazalis} Cazalis, J., Shah, T., Chai, Y., Jansen, K. and Kühn, S., 2024. Gaussian boson sampling for binary optimization. arXiv preprint arXiv:2412.14783.

\bibitem{Alcazar} Alcazar, J., Ghazi Vakili, M., Kalayci, C.B. and Perdomo-Ortiz, A., 2024. Enhancing combinatorial optimization with classical and quantum generative models. Nature Communications, 15(1), p.2761.

\bibitem{Fioroni} Fioroni, L. and Savona, V., 2025. Entanglement-assisted variational algorithm for discrete optimization problems. Communications Physics, 8(1), p.438.



\bibitem{Lucas} Lucas, A., 2014. Ising formulations of many NP problems. Frontiers in physics, 2, p.5.



\bibitem{Farhi} Farhi, E., Goldstone, J., Gutmann, S., Lapan, J., Lundgren, A. and Preda, D., 2001. A quantum adiabatic evolution algorithm applied to random instances of an NP-complete problem. Science, 292(5516), pp.472-475.

%==================paragraph 2

\bibitem{Wang} Rønnow, T.F., Wang, Z., Job, J., Boixo, S., Isakov, S.V., Wecker, D., Martinis, J.M., Lidar, D.A. and Troyer, M., 2014. Defining and detecting quantum speedup. science, 345(6195), pp.420-424.

\bibitem{Hauke-rpp-2020} Hauke, P., Katzgraber, H.G., Lechner, W., Nishimori, H. and Oliver, W.D., 2020. Perspectives of quantum annealing: Methods and implementations. Reports on Progress in Physics, 83(5), p.054401.




\bibitem{qxor-prl-2010} Jörg, T., Krzakala, F., Semerjian, G. and Zamponi, F., 2010. First-order transitions and the performance of quantum algorithms in random optimization problems. Physical review letters, 104(20), p.207206.

\bibitem{Young} Young, A.P., Knysh, S. and Smelyanskiy, V.N., 2010. First-order phase transition in the quantum adiabatic algorithm. Physical review letters, 104(2), p.020502.

\bibitem{qxor-pre-2011} Hen, I. and Young, A.P., 2011. Exponential complexity of the quantum adiabatic algorithm for certain satisfiability problems. Physical Review E—Statistical, Nonlinear, and Soft Matter Physics, 84(6), p.061152.

\bibitem{qxor-pra-2012} Farhi, E., Gosset, D., Hen, I., Sandvik, A.W., Shor, P., Young, A.P. and Zamponi, F., 2012. Performance of the quantum adiabatic algorithm on random instances of two optimization problems on regular hypergraphs. Physical Review A—Atomic, Molecular, and Optical Physics, 86(5), p.052334.

\bibitem{QC-prep-2013} Bapst, V., Foini, L., Krzakala, F., Semerjian, G. and Zamponi, F., 2013. The quantum adiabatic algorithm applied to random optimization problems: The quantum spin glass perspective. Physics Reports, 523(3), pp.127-205.

\bibitem{Brodoloni} Brodoloni, L., Astrakharchik, G.E., Giorgini, S. and Pilati, S., 2026. Energy gap of quantum spin glasses: a projection quantum Monte Carlo study. Physical Review Letters, 137(6), p.060403.

\bibitem{Damerow-prb-2026} Damerow, S. and Kehrein, S., 2026. Extension of the adiabatic theorem. Physical Review B, 113(16), p.165102.


\bibitem{Albash}	Albash, T. and Lidar, D.A., 2015. Decoherence in adiabatic quantum computation. Physical Review A, 91(6), p.062320.

\bibitem{Lidar-prl-2025} Munoz-Bauza, H. and Lidar, D., 2025. Scaling advantage in approximate optimization with quantum annealing. Physical Review Letters, 134(16), p.160601.
	
	
%==================paragraph 3

\bibitem{Muga-rmp-2019} Guéry-Odelin, D., Ruschhaupt, A., Kiely, A., Torrontegui, E., Martínez-Garaot, S. and Muga, J.G., 2019. Shortcuts to adiabaticity: Concepts, methods, and applications. Reviews of Modern Physics, 91(4), p.045001.

\bibitem{Polkovnikov} Čepaitė, I., Polkovnikov, A., Daley, A.J. and Duncan, C.W., 2023. Counterdiabatic optimized local driving. PRX Quantum, 4(1), p.010312.

\bibitem{Malla-prr-2024} Malla, R.K., Sukeno, H., Yu, H., Wei, T.C., Weichselbaum, A. and Konik, R.M., 2024. Feedback-based quantum algorithm inspired by counterdiabatic driving. Physical Review Research, 6(4), p.043068.



\bibitem{Marshall} Marshall, J., Venturelli, D., Hen, I. and Rieffel, E.G., 2019. Power of pausing: Advancing understanding of thermalization in experimental quantum annealers. Physical Review Applied, 11(4), p.044083.

\bibitem{Mehta-pra-2025} Mehta, V., De Raedt, H., Michielsen, K. and Jin, F., 2025. Unraveling reverse annealing: A study of D-Wave quantum annealers. Physical Review A, 112(1), p.012414.



\bibitem{Finin-srep-2020} Ayanzadeh, R., Halem, M. and Finin, T., 2020. Reinforcement quantum annealing: A hybrid quantum learning automata. Scientific reports, 10(1), p.7952.

\bibitem{ChenChen} Chen, Y.Q., Chen, Y., Lee, C.K., Zhang, S. and Hsieh, C.Y., 2022. Optimizing quantum annealing schedules with Monte Carlo tree search enhanced with neural networks. Nature Machine Intelligence, 4(3), pp.269-278.


\bibitem{Werner-arxiv-2026} Werner, M., Jonsson, M., García-Sáez, A., Riera, A. and Albash, T., 2026. Leveraging Landau-Zener-St\" uckelberg interference for accelerating diabatic quantum annealing. arXiv preprint arXiv:2606.09706.


%==================paragraph 4


\bibitem{Gui-comp-2006} Gui-Lu, L., 2006. General quantum interference principle and duality computer. Communications in Theoretical Physics, 45(5), pp.825-844.

\bibitem{Childs-arxiv-2012} Childs, A.M. and Wiebe, N., 2012. Hamiltonian simulation using linear combinations of unitary operations. arXiv preprint arXiv:1202.5822.

\bibitem{Berry-prl-2014} Berry, D.W., Childs, A.M., Cleve, R., Kothari, R. and Somma, R.D., 2015. Simulating Hamiltonian dynamics with a truncated Taylor series. Physical review letters, 114(9), p.090502.

\bibitem{Berry-ieee-2015} Berry, D.W., Childs, A.M. and Kothari, R., 2015, October. Hamiltonian simulation with nearly optimal dependence on all parameters. In 2015 IEEE 56th annual symposium on foundations of computer science (pp. 792-809). IEEE.

\bibitem{Qadd-prl-2016} Oszmaniec, M., Grudka, A., Horodecki, M. and Wójcik, A., 2016. Creating a superposition of unknown quantum states. Physical review letters, 116(11), p.110403.

\bibitem{Qadd-pra-2017} Li, K., Long, G., Katiyar, H., Xin, T., Feng, G., Lu, D. and Laflamme, R., 2017. Experimentally superposing two pure states with partial prior knowledge. Physical Review A, 95(2), p.022334.

\bibitem{Chatterjee-prl-2025} Chatterjee, A., Karthik, H.S., Mahesh, T.S. and Usha Devi, A.R., 2025. Extreme violations of Leggett-Garg inequalities for a system evolving under superposition of unitaries. Physical Review Letters, 135(22), p.220202.



\bibitem{Ramezanpour2018} Ramezanpour, A., 2018. Enhancing the efficiency of quantum annealing via reinforcement: A path-integral Monte Carlo simulation of the quantum reinforcement algorithm. Physical Review A, 98(6), p.062309.

\bibitem{Ramezanpour2022} Ramezanpour, A., 2022. Quantum walk in a reinforced free-energy landscape: Quantum annealing with reinforcement. Physical Review A, 106(1), p.012418.

\bibitem{Homayouni} Homayouni-Sangari, M. and Ramezanpour, A., 2026. Noise tolerance via reinforcement in the quantum search problem. Physical Review A, 114(2), p.022423.

%%%%%%%%%%%%%%%%%%%%%%%%%%%%%%%%%%%%%%%%%

\bibitem{Bapst-jstat-2012} Bapst, V. and Semerjian, G., 2012. On quantum mean-field models and their quantum annealing. Journal of Statistical Mechanics: Theory and Experiment, 2012(06), p.P06007.



\bibitem{grover-prl-1997} Grover, L.K., 1997. Quantum mechanics helps in searching for a needle in a haystack. Physical review letters, 79(2), p.325.


\bibitem{Roland-pra-2002} Roland, J. and Cerf, N.J., 2002. Quantum search by local adiabatic evolution. Physical Review A, 65(4), p.042308.

\bibitem{fdsearch-prl-2005} Grover, L.K., 2005. Fixed-point quantum search. Physical Review Letters, 95(15), p.150501.

\bibitem{fdsearch-pra-2017} Dalzell, A.M., Yoder, T.J. and Chuang, I.L., 2017. Fixed-point adiabatic quantum search. Physical Review A, 95(1), p.012311.















\end{thebibliography}
\end{document}